\documentclass[sigconf,nonacm]{acmart}
\AtBeginDocument{%
  }

\begin{document}

\title{Knowing when to stop: insights from ecology for building catalogues, collections, and corpora \vspace{12pt}}

\author{Jan Haji\v{c} jr.}
\authornote{Both authors contributed equally to this research.}
\email{hajicj@ufal.mff.cuni.cz}
\orcid{0000-0002-9207-567X}
\affiliation{
    \institution{Institute of Formal and Applied Linguistics \\  
    Charles University}
    \city{Prague}
    \country{Czech Republic \vspace{12pt}}
}

\author{Fabian C. Moss}
\authornotemark[1]
\email{fabian.moss@uni-wuerzburg.de}
\orcid{0000-0001-9377-2066}
\affiliation{%
\institution{Department of Musicology \\ Julius-Maximilians-Universität Würzburg}
  \city{Würzburg}
  \country{Germany \vspace{12pt}}
}

\renewcommand{\shortauthors}{}

\begin{abstract}
A major locus of musicological activity---increasingly in the digital domain---is the cataloguing of sources,
which requires large-scale and long-lasting research collaborations.  
Yet, 
the databases aiming at covering and representing musical repertoires are never quite complete, 
and scholars must contend with the question: how much
are we \textit{still} missing?
This question structurally resembles the `unseen species' problem in ecology, 
where the true number of species must be estimated from limited observations. 
In this case study, we apply for the first time the common Chao1 estimator to music, 
specifically to Gregorian chant. 
We find that, overall, upper bounds for repertoire coverage of the major chant genres range between 50 and 80\%.
As expected, we find that Mass Propers are covered better than the Divine Office, though not overwhelmingly so. 
%
However, the accumulation curve suggests that those bounds are not tight:
a stable $\sim$5\% of chants in sources indexed between 1993 and 2020 was new, so diminishing returns in terms of repertoire diversity are not yet to be expected. 
Our study demonstrates that these 
questions can be addressed empirically to inform musicological data-gathering, showing the potential of unseen species models in musicology.
\end{abstract}






\maketitle
\pagestyle{plain}

\section{Introduction}
\label{sec:introduction}

For decades, the humanities have been adding sources
to the digital domain, with massive attendant benefits for accessibility and cultural heritage preservation that 
enable cross-cutting research. With these large-scale efforts, music libraries, archives, and research institutions 
often pursue ambitious goals of accurately and/or completely representing a certain musical tradition or repertoire.
Immediately ensuing questions are: What is \textit{not} yet in the database? Can we know \textit{how much} we are still missing?  
We of course cannot answer these exhaustively without the prohibitive effort of actually cataloguing all extant sources, 
which in some cases is a virtually impossible endeavour as most sources are forever lost \cite{buringh2010manuscripts}. For instance, we can probably never reconstruct the tradition of medieval chant \textit{in its entirety}, regardless of our scholarly efforts. 


While one cannot have exact answers, however, one can aim for the next best thing: a principled estimate. 
There is hope in recognizing that historical musicologists, and historians more generally, share the problem of missing data with archaeologists, palaeontologists, and evolutionary biologists \cite{Gaddis2004_LandscapeHistoryHow}---and more recently also with data scientists.
In those fields, reliable methods and models for estimating how much one has not (yet) observed have been developed since 1943: the `Unseen Species' models \cite{fisher1943relation}. 
Recognizing structural analogies between the respective domains enables us to apply models developed in ecology and biology: we can use presently existing musicological databases to obtain credible estimates of the size of the lost repertoire. This has already proven useful in related areas, e.g. Shakespeare's poetic language~\cite{Efron1976_EstimatingNumberUnseen} and the study of medieval literature \cite{kestemont2022unseen}, which are direct inspirations for our work in this paper.  
%

Our main contribution is to apply, for the first time, unseen species models to music. 
For this case study, we choose the medieval European Gregorian chant repertoire as represented in the Cantus database. 
While there are numerous computational approaches to the analysis of chant melodies~\cite{cornelissen2020mode,cornelissen2020studying,helsen2021sticky,hornby2022analysis,Nakamura2023_HistoricalChangesModes,lanz2023,hajic2023,eipert2023monodikit,ballen2024}, 
the constitution of the repertoire itself 
has not yet been studied with formal models. 
Here, we estimate how much of this repertoire we have not yet seen and examine how it has accumulated.
We believe our foray into an `ecology of music' 
can lead to many more experiments and insights into the cultural ecosystem that is music throughout history.

\section{Case Study: Gregorian chant}
\label{sec:dataset}

In musicology, one of the most successful
and oldest digital cataloguing initiatives is the Cantus database of Gregorian chant \cite{lacoste2012cantus},\footnote{\url{https://cantusdatabase.org/}}
and the network of almost 20 interoperable databases,
interconnected via the Cantus Index network \cite{lacoste2022cantus}.
This massive undertaking has yielded a dataset of more than 
800,000 catalogue records across hundreds of sources. 
%
Given its exceptionally large scale,
it is tempting to treat Cantus as truly `big musicological data' that is 
representative of the entire tradition
of Gregorian chant, allowing scholars to adopt modern distant-reading methodologies \cite{Moretti2013_DistantReading,Rose2015_WritingBigData} and infer from it, amongst other things, distributional characteristics of this repertoire, such as its extent, shape, and diversity.
However, while the total number of catalogued chant \textit{melodies} is indeed huge, the number of catalogued chant \textit{manuscripts} (sources) is much smaller,
especially in relation to the estimated $\sim$30,000 extant chant manuscripts \cite{helsen2014omr}. 
Therefore, one has to be
careful in drawing conclusions based on statistical considerations only.

Here, we focus on the core chant repertoire for the Divine Office and for the Mass. This has the advantage that each chant is assigned a unique Cantus~ID (CID) that
unambiguously defines which catalogue record of a chant belongs to which ``chant species'' by merging variants of the same CID. As Gregorian chant is a tradition that does allow for this clarity, we adopt this crucial decision from the Cantus editors. 
Moreover, within this database, each catalogued source also has a unique identifier, allowing us to interpret sources as samples that contain a certain number of species (see Section~\ref{sec:formalisation}).

%

For the sake of simplicity and replicability, we use the established \texttt{CantusCorpus}~(v0.2) 
\cite{cornelissen2020studying}.\footnote{\url{https://github.com/bacor/cantuscorpus/releases/tag/v0.2}} 
While this dataset only reflects the state of the Cantus database up until 2020 and the additions from the last five years are absent, it is the most established dataset of chant, re-used in computational research \cite{cornelissen2020mode,lanz2023unsupervised,lanz2023}. 
%
%
Because different genres of chant
may vary differently, aggregating results across genres may obscure these differences in diversity between them. We therefore work with the genres separately, focusing on core office and mass propers genres.
The dataset used contains a total of 464532 chant catalogue records of the genres;
detailed statistics per genre
are provided in \autoref{tab:results}.


\section{Unseen Species Models on Gregorian Chant}
\label{sec:formalisation}

Musicological databases often aim to represent  entire repertoires, the tradition of European medieval chant in the case of Cantus.\footnote{For this first study we dispense of the fact that this repertoire---despite its canonical nature---is far from static, evolving both historically as well as geographically. This simplification does obviously not accurately represent the historical genesis of the repertoire, but it enables us to probe the usefulness of the model for this less complex scenario, and enables us study more complex cases in our future work. In the ecological analogy, it is as if the period over which we collected samples spanned several centuries.}
Translating cataloguing efforts to the context of unseen-species models means asking two related questions: 

\begin{enumerate}
    \item How many chants (unknown species) did we not find yet?
    \item If cataloguing continues, how quickly are we going to discover more repertoire?\footnote{This can also be rephrased as how many more sources we need to catalogue to likely see some target percentage of works from the complete tradition \cite{kestemont2022unseen}.}
\end{enumerate}

\noindent The first question is critical to the representativeness of Cantus. Having ``only'' indexed a few hundred
of the thousands extant manuscripts, how well is the entire tradition---at least in terms of repertoire---covered?
More formally, we are interested in estimating the overall number of different chants in the Gregorian repertoire (number of species; denoted by $S$ in what follows), drawing on the chants observed in extant catalogued sources (samples).
The second question then quantifies the expected `return on cataloguing investment' in terms of repertoire diversity.
%
%
%

\paragraph{Modeling the abundance of chants}
Some chants are very common (highly abundant; they occur in many manuscripts, perhaps also more than once per manuscript), while others are rare. The (unknown) probability to encounter a certain chant is thus proportional to the number of times that chant occurs in the whole repertoire. This is called the \textit{relative abundance} (or frequency) $p_i$ of chant $i$, and $\sum p_i = 1$, for $i=1,\dots, S$.\footnote{Relative abundances can be generalised further to include the probability of a species being detected \cite{chao2017deciphering}, but we focus on the simple case here.}
%
Suppose we inspect a sub-collection of sources (a sample) containing $n$ chants with (observed) abundances $X_i$ for chant $i$, so that

\begin{align}
    \sum_{i = 1 \dots S} X_i = n.
\end{align} 

\noindent
For some (possibly many) chants, we will have $X_i = 0$, which means that while they were at some point written down and sung by someone, they are not contained in the specific source we are looking at (they were not observed in the sample).
We define $f_r$ to be the absolute frequency of chants with abundance $r$:

\begin{align}
    f_r = \left|\{X_i\mid X_i=r\}\right|.
\end{align}

\noindent
So, $f_1$ is the number of chants observed once, $f_2$ is the number 
of chants observed twice, etc. We denote as $f_0$ the (unknown) number of chants in the repertoire that we have not seen at all. Then, the total number of distinct chants observed in the sample, $S_{\mathrm{obs}}$, is

\begin{align}
    S_{\mathrm{obs}} = \sum_{r = 1 \dots \infty} f_r,
\end{align} 

\noindent
which gives us an expression for our quantity of interest, the true size of the Gregorian repertoire, as:

\begin{align}
 S = S_{\mathrm{obs}} + f_0.   
\end{align}

\noindent
Since $S_{\mathrm{obs}}$ is known, the problem reduces to estimating $f_0$ from a sample. This is what Unseen Species models attempt to do.

\paragraph{Modeling chant incidence}
Instead of counting how many times each chant appeared in the dataset, we can also count \textit{incidence}, which is simply the presence or absence in a sample of sources, represented as a Boolean value (\textsc{True} or \textsc{False}). 
%
Instead of sampling $n$ catalogued chants, we sample $m$ chants and define $f_r$ to be the number of \textit{different} chants that appeared in exactly $r$ catalogued sources. 
That is, we only care about \textit{whether or not} a chant has appeared in a source. 

Preferring incidence follows naturally from the structure of chant data. Each written source is a sample, as though one were `trapping' what was sung at a certain place for a liturgical year. 
This implies that a chant recorded in only one source, even if used twice or more times, is still considered a singleton and contributes to $f_1$. 
The incidence-based approach thus asks: ``Was a given chant known and used there and then?'' rather than how often it was used. We believe this to be more appropriate: a chant being used in more than one liturgical position in a certain church should not necessarily imply the particular chant would be more likely to be used in other churches.\footnote{Abundance-based results do not lead to different conclusions on CantusCorpus.}

\paragraph{Chao estimators.}
The Chao1 estimator\footnote{We use the implementation of unseen species estimators from the \texttt{copia} library \cite{karsdorpIntroducingFunctionalDiversity2022}.} \cite{chao1984estimator} is one of many ways of estimating $S$ from $f_r, r > 0$.
It is formally defined as:
\begin{align}
    S = S_{\mathrm{obs}} + \frac{f_1^2}{2 f_2}
\end{align}

\noindent
Then, $f_0$ can be found simply as $S - S_{\mathrm{obs}}$, and \textit{species coverage} (the estimated ratio of observed species) as $c = S_{\mathrm{obs}} / S$.
The estimator only uses $f_1$ and $f_2$, the count of singletons (CIDs observed in one sample) and doubletons (observed in two);\footnote{When abundance and incidence data are both used, $f_r$ is usually denoted by $Q_r$ for incidence and the incidence-based estimator is called Chao2 \cite{chao1984estimator,chao2017deciphering}.} 
the intuition here is that the probability density reserved for unseen species should follow from how long the tail of observed incidences is.
We refer the reader to \cite{chao2017deciphering} for a constructive proof that follows from Good-Turing smoothing \cite{good1953smoothing,gale1995good}.
Importantly, the estimator is nonparametric: it can be used regardless of the underlying distribution of relative abundances or incidences $p_i$ \cite{chao1984estimator}.

The Chao estimators are lower bounds (see \cite{chao2017deciphering} for proof): the estimated $f_0$ is the \textit{minimum} expected number of unseen species (in our case: Cantus IDs). Conversely the estimated species coverage is an upper bound: a coverage of $0.5$ implies we have seen \textit{at most} half of all the chants in Gregorian repertoire.

We choose Chao over other estimators (Abundance Coverage Estimator --- ACE \cite{chao2004ace}, Jackknife \cite{smith1984jackknife}, or recent improvements on Good-Toulmin estimators \cite{orlitsky2016optimal,hao2020multiplicity}) because they are a conservative lower bound \cite{kestemont2022unseen}, thus minimising the chance of over-estimating the value of cataloguing further; because they have already been used in other applications of the unseen species model to the humanities \cite{karsdorpIntroducingFunctionalDiversity2022,kestemont2022unseen}; and because they are simple to interpret.

\setlength{\tabcolsep}{0.43em}
\begin{table}[]
    \centering
	\begin{tabular}{lrrrrrrrr}
	\toprule
\textbf{Genre} & \textbf{\# Chants} & \textbf{\# CIDs} & \textbf{\# sources} & \textbf{Chao1 $f_0$} & \textbf{max. $c$} \\
	\midrule
\textbf{\textit{Office}} & \textit{448041} & \textit{25805} & \textit{241} & \textit{20661} & \textit{0.56} \\    
\textbf{A}	& 205409 & 11158	& 231	& 8450 & 0.57 \\
\textbf{R}	& 102443 & 5099	& 213	& 4121 & 0.55 \\
\textbf{V}	& 94482 & 8163	& 214	& 8108 & 0.50 \\
\textbf{W}	& 35594 & 926	& 185	& 437 & 0.68 \\
\textbf{I}	& 10086 & 600	& 181	& 407 & 0.60 \\
	\midrule
\textbf{\textit{Mass Pr.}} & \textit{16511} & \textit{2268} & \textit{114} & \textit{999} & \textit{0.69} \\    
\textbf{In}	& 2025 & 207	& 50	& 154 & 0.57 \\
\textbf{InV}	& 1294 & 286	& 32	& 98 & 0.74 \\
\textbf{Gr}	& 2411 & 154	& 90	& 56 & 0.73 \\
\textbf{GrV}	& 1677 & 207	& 68	& 104 & 0.67 \\
\textbf{Al}	& 2382 & 405	& 73	& 244 & 0.62 \\
\textbf{AlV}	& 189 & 38	& 29	& 155 & 0.20 \\
\textbf{Of}	& 2209 & 158	& 43	& 37 & 0.81 \\
\textbf{OfV}	& 764 & 263	& 13	& 29 & 0.90 \\
\textbf{Cm}	& 2155 & 198	& 42	& 73 & 0.73 \\
\textbf{CmV}	& 253 & 154	& 4	& 686 & 0.18 \\
\textbf{Tc}	& 294 & 47	& 21	& 12 & 0.80 \\
\textbf{TcV}	& 858 & 203	& 21	& 37 & 0.85 \\

	\bottomrule
	\end{tabular}
	\caption{
    Unseen-species estimates for office genres (top rows) and mass propers (bottom rows), also with statistics of aggregated chants for the Office and the Proper of the Mass, respectively (italicised). We report the estimated lower bound on the number of unseen Cantus IDs (Chao1 $f_0$) and the implied maximum proportion of the repertoire in that genre already observed (max. coverage $c$).
    Repertoire statistics of the CantusCorpus~v0.2 dataset are given, broken down into individual genres.
    Genre abbreviations according to Cantus Index: Office chants (left), Mass Propers (right side). The table shows the number of chants, Cantus IDs (CIDs), and sources for each genre. Office chant genres: A = Antiphon, R = Responsory, V = Responsory Verse, W = Versicle, I = Invitatory antiphon; Mass Proper genres: In = Introit, Gr= Gradual, Al = Alleluia, Of = Offertory, Cm = Communion, Tc = Tract; appended V's refer to the corresponding verses.
    }\label{tab:results}
\end{table}


\section{Results and Discussions}
\label{sec:results}

\paragraph{Unseen species estimate.}
We now report the results from the incidence model applied to the \texttt{CantusCorpus} (v0.2) dataset, 
grouped by individual main chant genres for the Office and Mass Propers.
%
In \autoref{tab:results}, we report for each genre the number of Cantus IDs and sources it appears in, and then the unseen-species estimates:
the Chao1 lower bound on the number of unseen Cantus IDs of that genre, $f_0$, and the corresponding upper bound on the proportion of Cantus IDs of the genre that we have already seen, $S_{\mathrm{obs}}/(S_{\mathrm{obs}} + f_0)$.

As can be expected, the Mass Propers repertoire seems to be less diverse: already with much fewer sources, coverage is up to 0.69, while we have seen at most 0.56 of all Cantus IDs of the Office.

The genres considerably vary in coverage. Out of all the main mass genres, introits vary even more than alleluias, with approximately as much coverage as office antiphons. Tracts, on the other hand, are very consistent (again matching expectation). The comparison between tracts (Tc) and tract verses (TcV) also reveals the non-trivial properties of the Unseen Species model: while there are much fewer different Cantus IDs observed for tracts than the verses across the same number of sources, their coverage upper bound is somewhat lower than the much richer genre TcV.

The estimator reflects non-trivial phenomena in repertoire.
A constrained repertoire is found for offertory verses (OfV): though they have 263 different CIDs across only 13 sources, very few offertory verses are singletons. This is \textit{not} a sampling artifact of only indexing offertory verses in related sources, the sources come from all of Latin Europe and across multiple centuries. Rather, it reflects the disuse of OfV after the 13th century \cite[p.121]{hiley1993western}.
In contrast, second alleluia verses, found in 29 sources but with only 38 different Cantus IDs, and which one would thus trivially expect to be less diverse than OfV, seem radically more diverse. This reflects the unstable nature of Alleluias \cite[p.131]{hiley1993western} (and points towards an opportunity for deeper study \cite{hughes2005alleluia}).

\paragraph{Accumulation curve.}
The Chao1-derived upper bound on repertoire coverage tells us where we are in the cataloguing effort with respect to the total unobserved repertoire, but not where we are going.
This can be most succinctly characterised by the \textit{accumulation curve}, which shows the relationship between the number of catalogued chants and the number of distinct Cantus IDs observed overall. Accumulation curves are typically concave, representing `diminishing returns': as we catalogue more and more items, we expect the probability of seeing a new Cantus ID among the next $k$ observations to decrease, simply because there are fewer CIDs left to document. To determine how efforts devoted to further documenting the diversity of the chant repertoire will pay off, we would like to find out how far along this curve we are.

Fortunately, many sources in the Cantus database have a record of the year when they were catalogued (field \texttt{indexing\_date} in CantusCorpus). We can thus reconstruct the empirical species accumulation curve through time as Cantus grew between 1993 and 2020, and we can track how the probability that the next chant will require creating a new Cantus ID changed over time and project the rate at which new chants will be discovered in the future.

\begin{figure}
    \centering
    \includegraphics[width=1.02\linewidth]{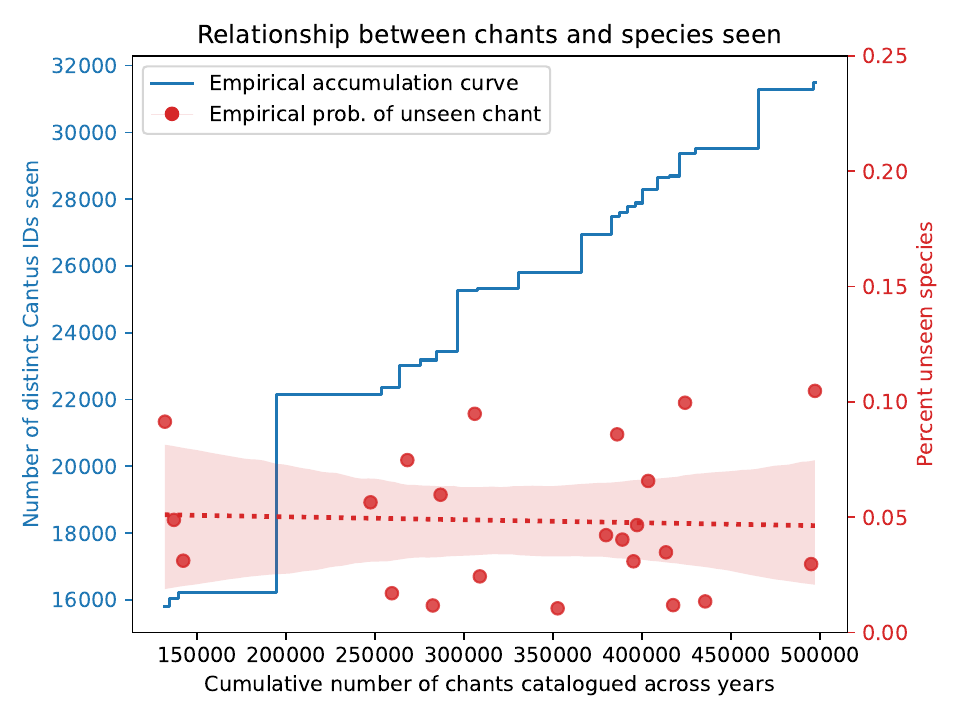}
    \caption{
    The empirical accumulation curve 
    (cumulative number of distinct CIDs; blue solid line)
    and the percentage of previously unseen CIDs for each year (red points) as new chants are added. 
    Linear regression (red dotted line; 95\% confidence interval plotted) shows that the probability of observing an unseen CID does not change as more chants are catalogued. 
    This implies the empirical accumulation curve is approximately linear: there are no ``diminishing returns''. 
    The x-axis is the total number of indexed chant records through these years. 
    Note that a different number of chants was catalogued each year, so time does not progress linearly along the x-axis; each step of the blue empirical accumulation curve and each red dot correspond to a year.}
    \label{fig:enter-label}
\end{figure}

The empirical accumulation curve over chants ordered into blocks by their indexing years is shown in \autoref{fig:enter-label} (blue graph), together with the probabilities of encountering a new Cantus ID when cataloguing a chant (red graph), shown per batch for each year of cataloguing year. Linear regression over the probabilities of observing unseen Cantus IDs\footnote{Implemented with \texttt{scipy.stats.linregress}. Linear regression can be applied, rather than a weaker nonparametric model, because the proportions of new Cantus IDs per year are normally distributed ($p=0.21$, \texttt{scipy.stats.normaltest}).} indicates there is no reason to think the probability of observing an unseen chant is not constant in time ($p > 0.9$ for the null hypothesis that the slope is $0$), with the intercept at approx. $P(\text{unseen}) = 0.05$. 

This implies that the empirical accumulation curve between 1993 and 2020 is approximately linear, and the Cantus cataloguing effort is not likely to face diminishing returns in chant repertoire diversity in the near future. (Also, the rate of discovery of new CIDs does not increase.) Note that due to loss of manuscripts, it is possible we will not get out of the approximately linear accumulation curve regime even if we do catalogue those 30,000 extant sources, if the true medieval diversity was orders of magnitude greater.

\vspace{48pt}

\section{Conclusions}
\label{sec:conclusions}

Can music librarians or archivists know whether continuing collection efforts will yield new discoveries? Should resources rather be redirected elsewhere? 
We propose to estimate how much remains to be found using Unseen Species models from ecology. Exemplifying the approach using the Cantus database of Gregorian chant,
%
we show that the cataloguing efforts of Cantus are nowhere close to the point of diminishing returns: 
it is not yet time to stop! 

At the same time, this outcome needs to be taken with caution, as some of the decisions and assumptions that went into the model are somewhat simplistic. For instance, the geographical provenance of sources is not at all taken into account.
Phylogenetic diversity metrics may help incorporate such factors~\cite{morlon2010phylogenetic,chao2014unifying}.

The implications of introducing methodologies from ecology and cultural evolution to (digital) musicology and archival studies are much broader, and we aim to explore further avenues in our future work.
%
%
For chant, besides the relationship between diversity and provenance, one could consider later layers of repertoire (for example tropes \cite{eipert2025corpus}), and more generally study diversification across time. 
Also, we could try to designate small subsets of liturgical positions that are good predictors of expected diversity in a particular source, providing guidance for individual cataloguing decisions.
Naturally, there are ample applications beyond chant scholarship.  
Using RISM~\cite{Hankinson2024_NavigatingRISMData} we could estimate how much remains to be discovered from different composers.  
Folk music databases \cite{Carvalho2023_ComputationalSimilarityPortuguese,Schaffrath1995_EssenFolksongDatabase,Janssen2018_RetainedLostTransmission} as well as broader ethnomusicological data \cite{apjok2024polyphony,Lisniak2025} likewise are promising data troves guiding cultural heritage conservation.\footnote{Note that methods of estimating biodiversity help allocate resources for conservation of natural heritage. These usually involve an additional phylogenetic component in the estimation.}

We are convinced that exploring further methodological connections between ecology and musicology will enable us to learn more about the `biodiversity' of other (digital) libraries and repertoires.
While, oftentimes, digital libraries are merely viewed as data providers for computational analysis, we hope to have emphasized here that computational models can also feed back into and enrich all sorts of collection processes of musicological data---catalogues, collections, and corpora.

\vspace{72pt}

\begin{acks}
This work was supported by the Social Sciences and Humanities Research Council of Canada by the grant no. 895-2023-1002, Digital Analysis of Chant Transmission, and the project ``Human-centred AI for a Sustainable and Adaptive Society'' (reg. no.: CZ.02.01.01/00/ 23\_025/0008691), co-funded by the European Union. The computing infrastructure is provided by the LINDAT/CLARIAH-CZ Research Infrastructure (https://lindat.cz), supported by the Ministry of Education, Youth and Sports of the Czech Republic (Project No. LM2023062).
\end{acks}

\vspace{0pt}

\bibliographystyle{ACM-Reference-Format}
\bibliography{aaa_bibliography}










\end{document}